\UseRawInputEncoding
\documentclass[prb,reprint,amsmath,amssymb,showpacs,superscriptaddress,floatfix,longbibliography]{revtex4-1}
\usepackage[breaklinks=true,colorlinks,citecolor=blue,linkcolor=blue,urlcolor=blue]{hyperref}
\usepackage[T1]{fontenc}
\usepackage{epsfig,mathrsfs,color,latexsym,subfigure,marginnote,gensymb}
\usepackage{graphicx}
\usepackage{float}
\usepackage{makecell}
\usepackage{textalpha}
\usepackage[normalem]{ulem}
\usepackage{hyperref}
\usepackage{tikz}
\usepackage{array}\usepackage{makecell}
\newcolumntype{?}{!{\vrule width 1pt}}
\renewcommand{\BibitemShut}[1]{}
\makeatletter
\newcommand*{\addFileDependency}[1]{
  \typeout{(#1)}
  \@addtofilelist{#1}
  \IfFileExists{#1}{}{\typeout{No file #1.}}
}
\makeatother

\begin{document}
\title{Accelerating microstructure modelling via machine learning: a new method combining Autoencoder and ConvLSTM}
\author{Owais Ahmad}
\email{owaisah@iitk.ac.in}
\author{Naveen Kumar}
\email{knaveen@iitk.ac.in}
\author{Rajdip Mukherjee}
\email{rajdipm@iitk.ac.in}
\author{Somnath Bhowmick}
\email{bsomnath@iitk.ac.in}
\affiliation{Department of Materials Science and Engineering, Indian Institute of Technology Kanpur, Kanpur 208016, India}

\date{\today}
\begin{abstract}
Phase-field modeling is an elegant and versatile computation tool to predict microstructure evolution in materials in the mesoscale regime. However, these simulations require rigorous numerical solutions of differential equations, which are accurate but computationally expensive. To overcome this difficulty, we combine two popular machine learning techniques, autoencoder and convolutional long short-term memory (ConvLSTM), to accelerate the study of microstructural evolution without compromising the resolution of the microstructural representation. After training with phase-field generated microstructures of ten known compositions, the model can accurately predict the microstructure for the future $n^{th}$ frames based on previous $m$ frames for an unknown composition. Replacing $n$ phase-field steps with machine-learned microstructures can significantly accelerate the in silico study of microstructure evolution.

\end{abstract}
\maketitle
\section{Introduction}
\label{intro}
In recent years, the rise of artificial intelligence (AI) in science and technology has been phenomenal. With the development of sophisticated machine learning algorithms and the availability of vast amounts of data, artificial intelligence has become an indispensable tool for solving complex problems in various fields. AI has also demonstrated great promise in materials science. One of the most active research domains is analyzing vast amounts of data by machine learning algorithms for the accelerated discovery of new materials.\cite{XIONG2020109203,jennings2019genetic} AI algorithms have been trained for microstructure analysis,\cite{DECOST2015126,holm2020overview} additive manufacturing,\cite{HERRIOTT2020109599} mapping materials properties to atomic-scale imaging,\cite{han2022materials} and diagnosing materials failure before they occur to reduce downtime and increase productivity.\cite{LI2021109726}

The phase-field method is a powerful computational tool to model and study microstructure evolution and related properties, including solidification,\cite{chatterjee2008phase, HOTZER2015194, ZHAO20191044} precipitate growth, \cite{mukherjee2009, Mukhrjee2010} grain growth, \cite{PhysRevLett.86.842, CHANG201767, verma_mukherjee} coarsening,\cite{MOLNAR20126961} effect of external field,\cite{Gururajan2007, CHAFLE} and spinodal decomposition.\cite{bhattacharyya2003study,ramanarayan2003spinodal} Apart from materials science, it has applications in various domains.~\cite{doi:10.1142/S0218202511500138, GARCKE2021103192, PhysRevE.69.021603} Since the microstructural images in phase-field models are represented by a system of continuously evolving variables in the spatial and temporal domain, this kind of fidelitous phase-field models demand a discretized spatiotemporal representation by partial differential equations, making their implementation computationally expensive and cumbersome, which motivated the researchers to minimize the computational costs by primarily leveraging advanced numerical methods,\cite{SEOL20035173, Muranushi_2012, jiao2021designing} and high-performance computational architectures.\cite{hunter2011large,vondrous2014parallel,miyoshi2017ultra}

\begin{figure*}
\includegraphics[width=2\columnwidth]{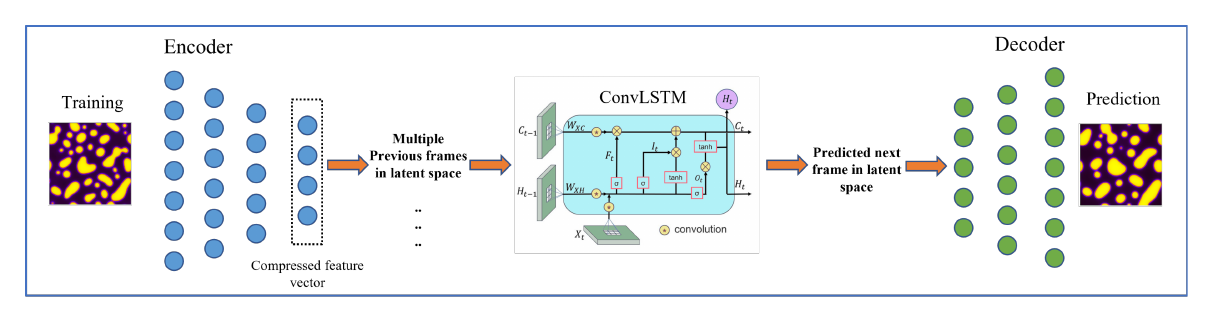}
\caption{Workflow of training machine learning model with phase-field generated microstructures and building a machine-learned surrogate model for accelerated prediction of microstructure evolution.}
\label{fig1}
\end{figure*}

As machine learning and deep learning have touched every domain, microstructure evolution is no exception. Researchers have already started leveraging artificial intelligence's power to do faster simulations.\cite{,hu2022accelerating} In this work, we propose a new method for accelerated prediction of microstructure evolution via a machine-learned surrogate model. Fig.~\ref{fig1} summarizes the workflow. Similar to the earlier works\cite{montes2021accelerating,hu2022accelerating} initially, we generate a dataset of 1000 images each for ten different compositions ranging from $c_{avg}=0.25$ to $0.5$  with the help of phase-field calculations. The size of each frame is $256\times 256\times 3$, where the last digit represents the number of channels (red, green, and blue) in the image. Since the dimensions are huge for 10000 images ($256\times 256\times 3$, i.e., 196608 per image), the required computational resource is also very high and time-consuming. To overcome this problem, we need to apply some dimensionality reduction techniques. In this work, we deploy the autoencoder method.~\cite{doi:10.1126/science.1127647,WANG2016232} The encoder part reduces the dimensions to 32x32x8 (i.e., 8192), and using this transformed version of data in latent space, we train the model to learn the spatial and temporal variation in the dataset. For predicting time series image data, we create a model with convolutional long short-time memory (ConvLSTM).\cite{NIPS2015_07563a3f} The model can accurately predict the microstructure for the future $n^{th}$ frames based on previous $m$ frames for an unknown composition. Finally, the decoder takes this predicted frame (still in latent space) and projects it into its original dimensions. Combined with the autoencoder, ConvLSTM proves to be a very robust technique to learn the spatiotemporal variation of microstructure with a minimal dataset for the binary phase. Our method is comparable to the ones carried out using principal component analysis (PCA) with recurrent neural network (RNN) and PCA with LSTM.\cite{hu2022accelerating,montes2021accelerating} 

The paper is organized in the following manner: In section~\ref{pfm}, we describe the details of the Phase-field model and data generation. In section~\ref{encoder}, we discuss the autoencoder method for dimensionality reduction. In section~\ref{convlstm}, we talk about spatiotemporal prediction using convLSTM, then reconstructing the microstructure using the decoder. In section~\ref{discussion}, we discuss the salient features of the method used in this work and compare it with other possible techniques. Finally, we conclude the paper in section~\ref{conclusion}.

\section{Phase-field model for spinoidal decomposition in a binary alloy}
\label{pfm}
\begin{figure}
\includegraphics[width=\columnwidth]{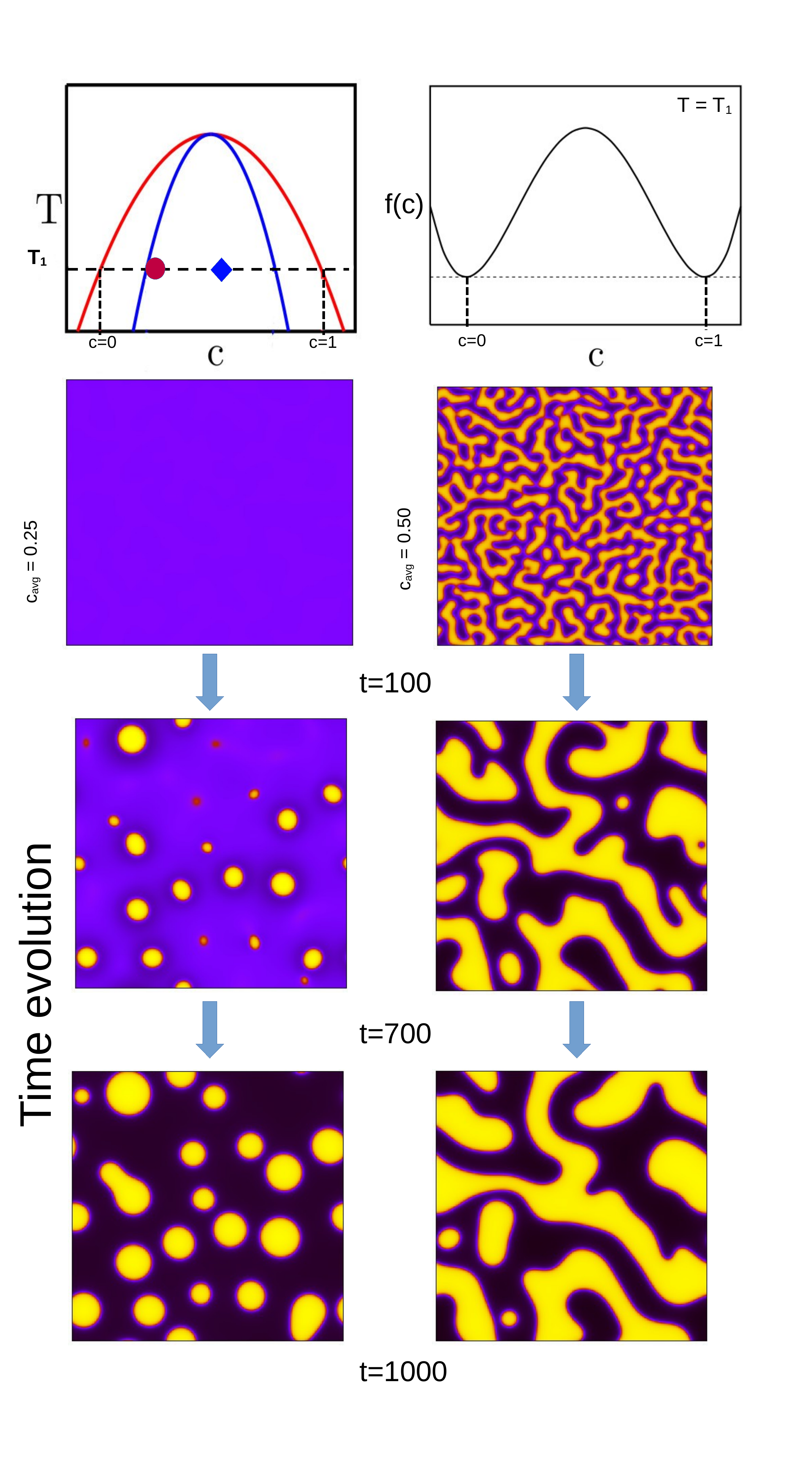}
\caption{Figure representing a  phase diagram with miscibility gap (top left) and the corresponding bulk free energy $f(c)$ versus composition $c$ diagram at temperature $T=T1$ (top right). The micorstructures show the time evolution (spinodal decomposition) during isothermal ($T=T1$) aging of alloys with two different initial compositions $0.25$ and $0.5$, respectively,  at time $t=100,700$ and $1000$. Red (circle) and blue (diamond) points on the phase diagram show these two compositions, respectively. Total ten thousand such images (ten different compositions , one thousand time frames per composition) are used as training set.}
\label{fig2}
\end{figure}
\begin{figure}
\includegraphics[width=\columnwidth]{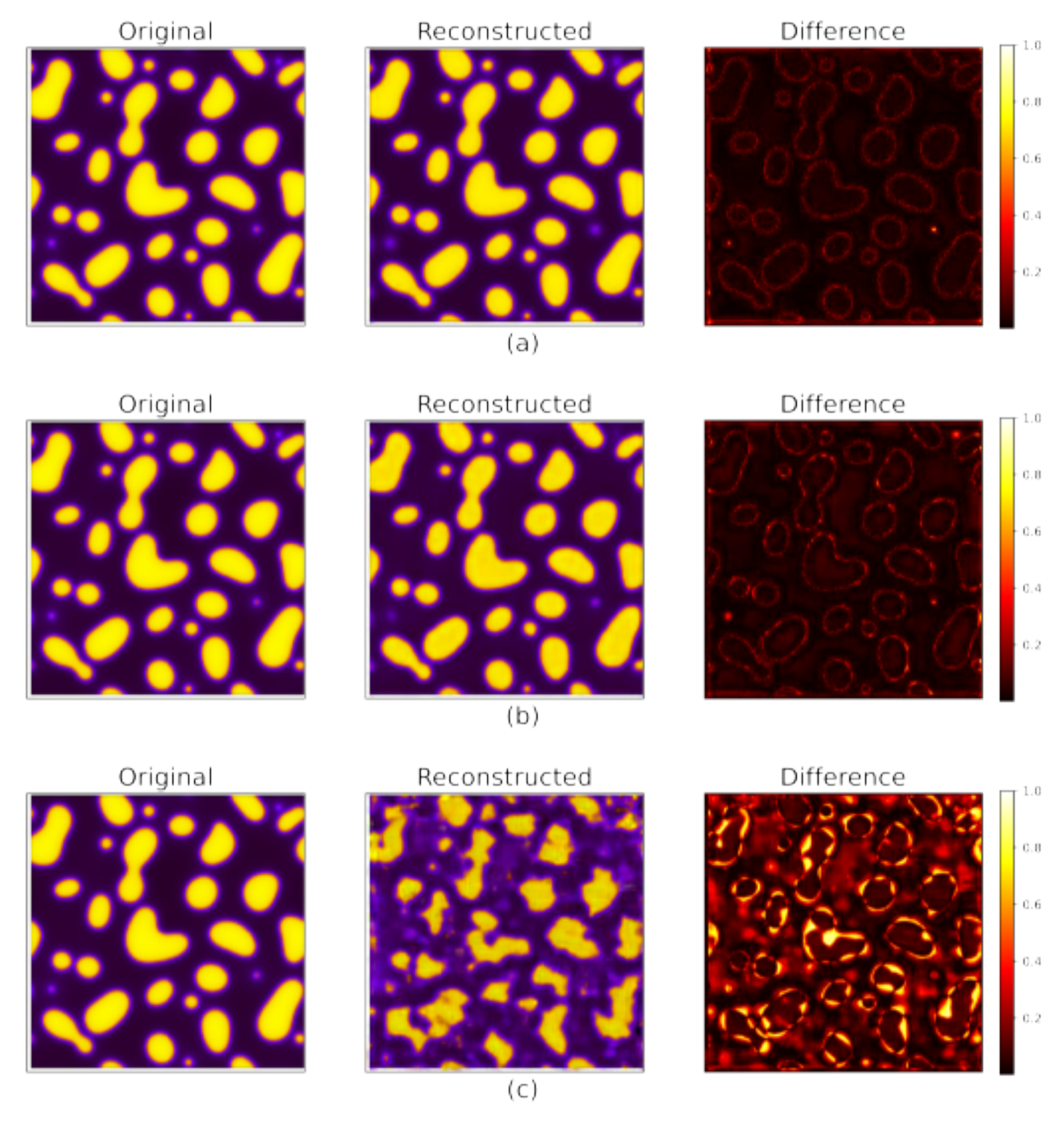}
\caption{Images are reduced in dimensions using the autoencoder. We find that 2000 images are sufficient for training the autoencoder; a heat map comparison of the original (encoded) and reconstructed (decoded) images using (a) 2 cells, (b) 3 cells, and (c) 4 cells autoencoders is presented. Reconstructed images are comparable to the original ones when two to three cells are used.}
\label{fig3}
\end{figure}
The present study employs a phase-field model to generate a training dataset of microstructure evolution during spinodal decomposition in a $A-B$ binary alloy. A schematic phase diagram is shown in Fig.~\ref{fig2} (top left), which shows the chemical spinodal lines (in blue) and the miscibility gap (in red), respectively. The total free energy of the alloy is ~\cite{doi:10.1063/1.1744102}, 
\begin{equation}
F=\int_V[f(c)+\kappa(\nabla c)^2] dV.
\label{tot_free_energy}
\end{equation}
The local composition of the system is denoted by a conserved phase-field variable $c(\boldsymbol{r},t)$ (space: $\boldsymbol{r}$ and time: $t$). The bulk free energy density $f(c)$ is schematically shown in Fig.~\ref{fig2} (top right), represented by a double well potential and is given by, 
\begin{equation}
f(c)=W c^2 (1-c)^2,
\label{free_energy_density}
\end{equation}
where $W$ is a constant determining the potential barrier height between the two equilibrium phases corresponding to the compositions  $c=0$ and $c=1$, respectively. Spinodal decomposition occurs in the composition range between the two inflection points where $\partial^2f/\partial c^2<0$.  The gradients in local composition also contribute to the total free energy as given by the term $\kappa (\nabla c)^2$ in Eq.~\ref{tot_free_energy}, where $\kappa$ is the gradient energy coefficient. The spatiotemporal evolution of the conserved phase-field  variable (composition $c$, in this case) is governed by the Cahn-Hilliard equation~\cite{cahn1961spinodal}: 
\begin{align}
\dfrac{\partial c}{\partial t}=M \left[\nabla^2 g(c)-2\kappa\nabla^4 c\right],
\label{CH}
\end{align}
where $M$ is the atomic mobility (assumed to be a constant), and $g(c)=\partial f/\partial c$. Further details about the Phase-field model and its numerical implementation are provided in the Supplemental Material.

Using Latin hypercube sampling, we generate ten combinations of phase fractions values, $\phi_A$, and $\phi_B$. Since we are interested in studying the spinodal decomposition of a binary system, the $\phi_A$ value should lie within a range of 0.25 and 0.75. However, because of the symmetry of the potential, the minimum and maximum values of $\phi_A$ are set to 0.25 and 0.5, respectively. The phase mobilities for both components are set to 1. The simulations are performed on a 2D square domain, discretized with $256 \times 256$ grid points, and the microstructure evolution and growth are allowed for 1000-time steps.  Fig.~\ref{fig2} shows the microstructure evolution in two different alloys, having initial compositions $0.25$ and $0.5$, respectively. We save the information of the microstructural state every single time step, yielding a total of 1000 time-frames per composition. Since there are ten compositions, we have a total of 10000 microstructure evolution images of $256 \times 256$ resolution. The initial compositional field is distributed arbitrarily in space, and the microstructure has no discernible features from frame $t_{0}$ to $t_{20}$. Thus, we discard the initial part and initiate training at frame $t_{20}$. The subdomains develop rapidly between frames $t_{20}$ and  $t_{100}$, followed by a smooth and consistent agglomerating outgrowth of the microstructure from frames $t_{100}$ to $t_{1000}$. We want our machine learning model to be able to anticipate both the rapid development and gradual growth phases of microstructure evolution. We evaluate the model using 80-20 train-test splits and verified that the outcome is similar for 60-40 and 70-30.

\section{Dimensionality reduction with Encoder and reconstruction with Decoder}
\label{encoder}
The enormous dimensionality of the phase-field data in the format of the microstructural image is precisely where the manifold hypothesis~\cite{Bostanabad2018} can be leveraged to establish an accelerated framework for studying the microstructural evolution. However, for the microstructure-learning model to operate effectively, one must transform the $256\times 256\times 3$ phase-field data per frame via a dimensionality-reduction procedure into a more compact and manageable dataset. A dimensionality-reduction algorithm aims to describe the data with fewer characteristics while retaining as much information as possible. This work uses the autoencoder for dimensionality reduction and transforming phase-field microstructure data in a smaller latent space for the convolutional long short-term memory (ConvLSTM) model to learn more efficiently. The performance of another popular dimensionality reduction method, the principal component analysis (PCA),\cite{abdi2010} is compared with the autoencoder, and the latter is found to be more accurate.

PCA is a linear transformation of the high-dimensional data that discards the insignificant modes (eigen/singular) with lower eigen/singular values, transforming the data into a low-dimensional form. Due to the non-linear nature of the system, employing PCA to decrease the dimension of the microstructure representation may result in the loss of vital information if only a few appropriate principle components are examined. We need a non-linear mapping technique from a high-dimensional spatial version to a low-dimensional latent space while avoiding loss of information, as in the case of principal component analysis (PCA). An autoencoder does precisely this. The encoder reduces the dimension to lower latent dimensions via a non-linear mapped version of high-dimensional microstructure data $\rm{(p, q, t)}$ into a low-dimensional version but in latent space represented by $(r)$. At the same time, the decoder learns the reverse mapping from low-dimensional latent space to high-dimensional microstructure. Mathematically, we can write this as,
\begin{equation}  
\label{eqencode}
\alpha_{\theta_{enc}} \colon \phi{\rm{({p},{q},{t})}} \rightarrow \tilde{\phi}({r}),
\end{equation}
\begin{equation}  
\beta_{\theta_{dec}} \colon \tilde{\phi}({r}) \rightarrow \phi{\rm{({p},{q},{t})}},
\end{equation}  
here $\alpha$ and $\beta$ represent the mapped versions as transformed by the encoder and the decoder, respectively. In Eq.~\ref{eqencode}, the encoder takes $\phi$ (p,q,t)  $\in$ $\mathbb{R}^{256\times 256\times 3}$ as input and maps it to a latent space $\tilde{\phi}({r})$ $\in$ $\mathbb{R}^{l_d}$ with $l_d$ dimensions. The error $\mathcal{L}_{ae}$ given below is minimised while training autoencoder 
\begin{equation} 
\mathcal{L}_{ae} = \underset{\theta_{ae} = (\theta_{enc},\theta_{dec})}{min} \left\| \phi(p,q,t) - \tilde{\phi}(p,q,t;\theta_{ae}) \right\|^{2}_{2}.
\end{equation}
Here $\theta_{ae}$ represents the parameters for training autoencoder. This transformed data in latent space is the training data for the ConvLSTM model. We observe a faster convergence in training ConvLSTM while feeding the dimensionally reduced data compared to actual data. The predicted output from ConvLSTM is then transformed back to its original dimensions (in the form of a microstructure) with the decoder, the second part of the autoencoder. 

The optimization of the autoencoder (both encoder and decoder) architecture is initially undertaken to preserve the majority of the features in the latent space. A series of tests are conducted, starting with adding two cells and gradually increasing to four. As depicted in Fig.~\ref{fig3}, it is observed that architectures featuring three or fewer encoder layers performed much better compared to four and above encoder layers. The transformed data loses its correlation with the features in more than three-layer encoders, and the decoder cannot accurately reconstruct the data from the transformed features. In the first row of Fig.~\ref{fig3}, two cells are taken for encoding; the reconstructed image is very close to the original image, and the heat map shows tiny red regions, implying very little difference between the original and reconstructed image. In the second row, Fig.~\ref{fig3}, three cells are taken for encoding; the reconstructed image is close to the original image but not as good as the case of two cells. As we increase the number of cells to four in the third row of Fig.~\ref{fig3}, the reconstructed image gets distorted, and the difference in certain regions reaches yellow in the heatmap, implying a significant difference in actual and reconstructed pixel values. Finally, we choose to use a three-cell encoder, as it offers a higher dimensionality reduction with minimal data loss.

In order to optimize the speed of the model training, experimentation is conducted using a different number of images in the dataset. The model is trained using datasets of 500, 1000, 2000, 3000, 4000, and 5000 images for 1000 epochs. The results indicate that while the architecture can learn the microstructure evolution profile using 500 images, it struggles to track the phases within that profile accurately. However, as the number of images in the training dataset increases, the model performs with an accuracy of 98\%. However, its performance in terms of loss and accuracy shows minimal improvement after 2000 images. 

\begin{figure}
\includegraphics[width=0.8\columnwidth]{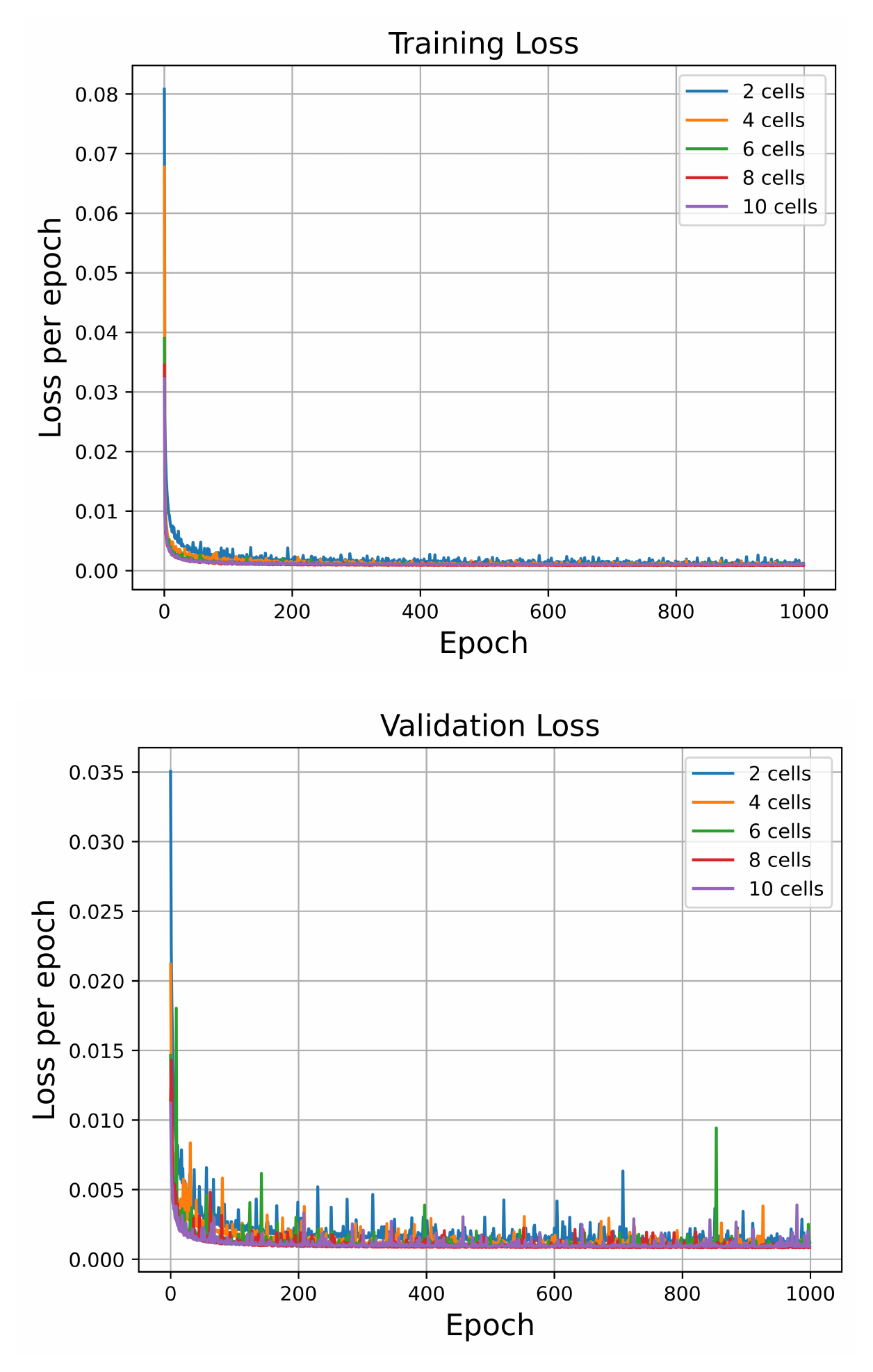}
\caption{The ConvLSTM model is trained with a compressed image dataset in latent space. The architecture of the ConvLSTM model comprises 2, 4, 6, 8, and 10 cells. The illustration compares training and validation loss for different ConvLSTM architectures.}
\label{fig4}
\end{figure}

\section{Training ConvLSTM, prediction and reconstructing microstructure with decoder}
\label{convlstm}
Convolutional long short-term memory (ConvLSTM) is a neural network architecture that combines the advantages of convolutional neural network (CNN)~\cite{computation11030052} with LSTM networks.~\cite{NIPS2015_07563a3f} The ConvLSTM architecture can analyze and learn spatial information with its temporal dependencies, such as in video or time series data. Inputs $\mathcal{X}_1$, $\mathcal{X}_2$, ... $\mathcal{X}_t$, cell outputs $\mathcal{C}_1$, $\mathcal{C}_2$ ... $\mathcal{C}_t$, hidden states $\mathcal{H}_1$, $\mathcal{H}_2$ ... $\mathcal{H}_t$ and gates $i_t$,$f_t$,$o_t$ all of these in case of ConvLSTM are 3D tensors, the last two dimensions being spatial. This is the advantage of ConvLSTM over LSTM because the spatial information is lost in the latter. Following equations represent the fundamental structure of a ConvLSTM cell. Here, \lq*\rq{} represents the convolution operator, and \lq$\circ$\rq{} represents the Hadamard product.
\begin{equation} 
i_t = \sigma(W_{xi} * \mathcal{X}_t + W_{hi} * \mathcal{H}_{t-1} + W_{ci} \circ \mathcal{C}_{t-1} + b_i)
\end{equation}
\begin{equation} 
f_t = \sigma(W_{xf} * \mathcal{X}_t + W_{hf} * \mathcal{H}_{t-1} + W_{cf} \circ \mathcal{C}_{t-1} + b_f)
\end{equation}
\begin{equation} 
\mathcal{C}_t = f_t \circ \mathcal{C}_(t-1) * i_t \circ tanh(W_{xc}*\mathcal{X}_t + W_{hc} * \mathcal{H}_{t-1} + b_c)
\end{equation}
\begin{equation} 
o_t = \sigma(W_{xo} * \mathcal{X}_t + W_{ho} * \mathcal{H}_{t-1} + W_{co} \circ \mathcal{C}_t + b_o)
\end{equation}
\begin{equation} 
\mathcal{H} = o_t \circ tanh(\mathcal{C}_t)
\end{equation}
Here $W_{xi}$, $W_{hi}$, $W_{ci}$,$W_{xf}$, $W_{hf}$, $W_{cf}$, $W_{xc}$, $W_{hc}$, $W_{xo}$,$W_{ho}$ and $W_{co}$ represents weights for respective variables while $b_i$, $b_f$, $b_c$, and $b_o$ represent the bias for each gate. The ConvLSTM predicts the future state of a grid cell based on the inputs and previous states of its immediate neighbors. Utilizing a convolution operator for the state-to-state and also with input-to-state transitions makes this simple to implement. Before conducting the convolution operation, padding is required to ensure that the output states and the inputs must have the same number of rows and columns. Using the state of the outer environment for computational purposes might be interpreted as padding for the hidden states at the boundary points. Before the first input, we typically set the LSTM states to zero, eliminating the dependency on the future. If we set padding as zero on the hidden states, it would set the state of the outside world to zero, assuming that it is unaware of the outside world. As the microstructures can be periodic, the padding has been set as same for this work.
\begin{figure}
\includegraphics[width=\columnwidth]{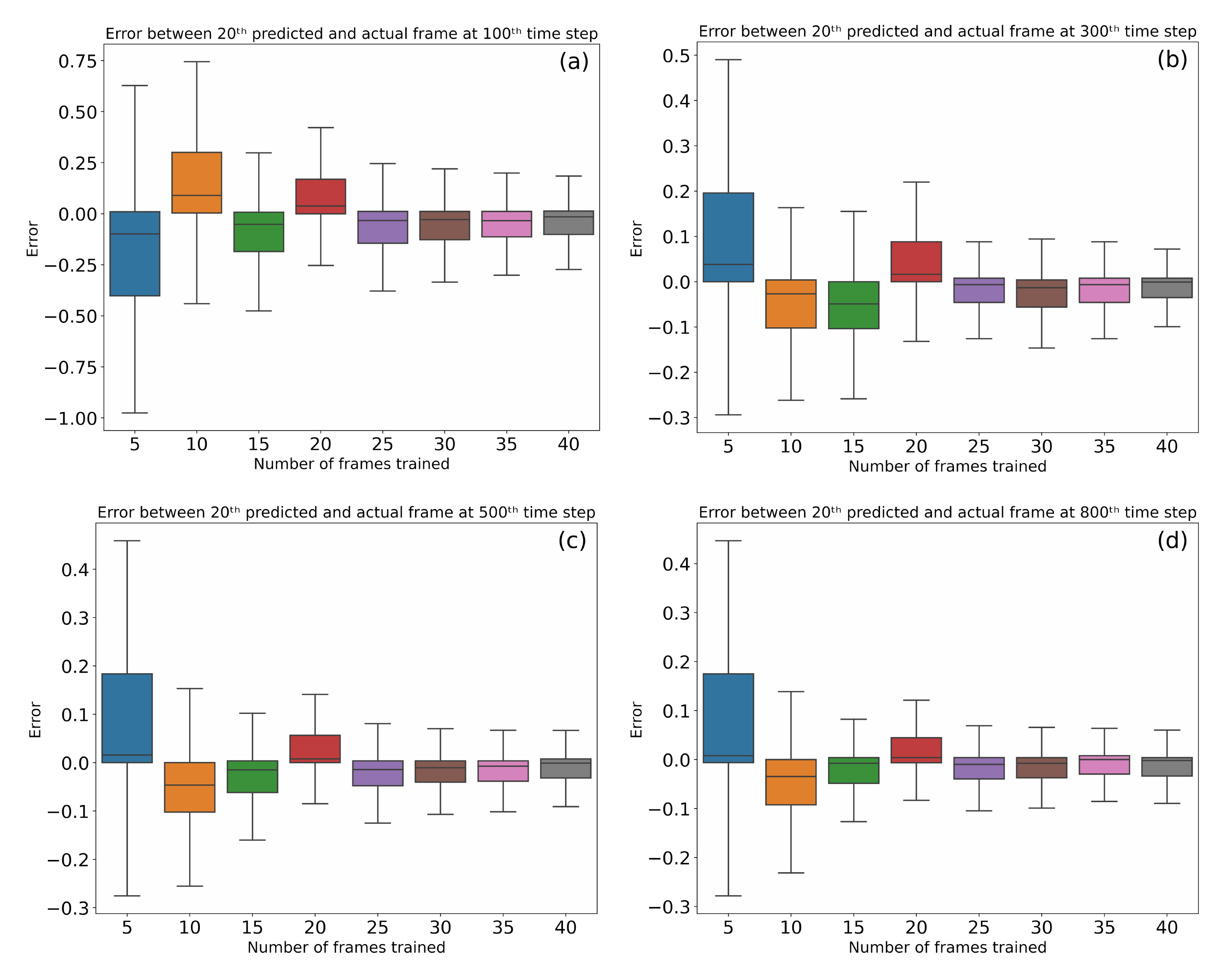}
\caption{This graph compares the relative error between the actual and predicted frame. For predicting the 20$^{th}$ frame, the number of previous frames used are 5, 10, 15, 20, 25, 30, 35, and 40. The comparison is made at different stages of the microstructure evolution: (a) 120$^{th}$, (b) 320$^{th}$, (c) 520$^{th}$, and (d) 820$^{th}$ time steps.}
\label{fig5}
\end{figure}

After reducing the microstructure dataset to latent dimensions using the optimized autoencoder model, the ConvLSTM architecture is applied to learn the changes in the microstructure with respect to time. In the process of optimizing the ConvLSTM architecture, the number of cells is gradually increased in steps of two as 2, 4, 6, 8, and 10. Fig.~\ref{fig4} compares the training loss and the validation loss for the different numbers of cells. The loss for ConvLSTM training is defined as the mean squared error(MSE),
\begin{equation}
\operatorname{MSE}_{a_j}=\frac{1}{K N} \sum_{k=1}^K \sum_{i=1}^N\left(\hat{a}_j^{(k)}\left(t_i\right)-\tilde{a}_j^{(k)}\left(t_i\right)\right)^2.
\end{equation}
Here $N$ is the number of time frames for which the error is calculated, and $K$ represents the total number of microstructure evolution predictions for which the error is calculated. $\hat{a}_j^{(k)}$ and $\tilde{a}_j^{(k)}$ represent the actual and predicted value of the pixel in latent space, respectively. We find that two and four ConvLSTM cells outperform all other architectural configurations when trained and compared over 1000 epochs. Since four ConsLSTM cells offer minimum loss at the expense of minimum computational cost, we use the same for the rest of the work. Though 100 epochs are sufficient for making the predictions, we show up to 1000 epochs to illustrate that model stays stable over 1000 epochs.

\begin{figure*}
\includegraphics[width=2\columnwidth]{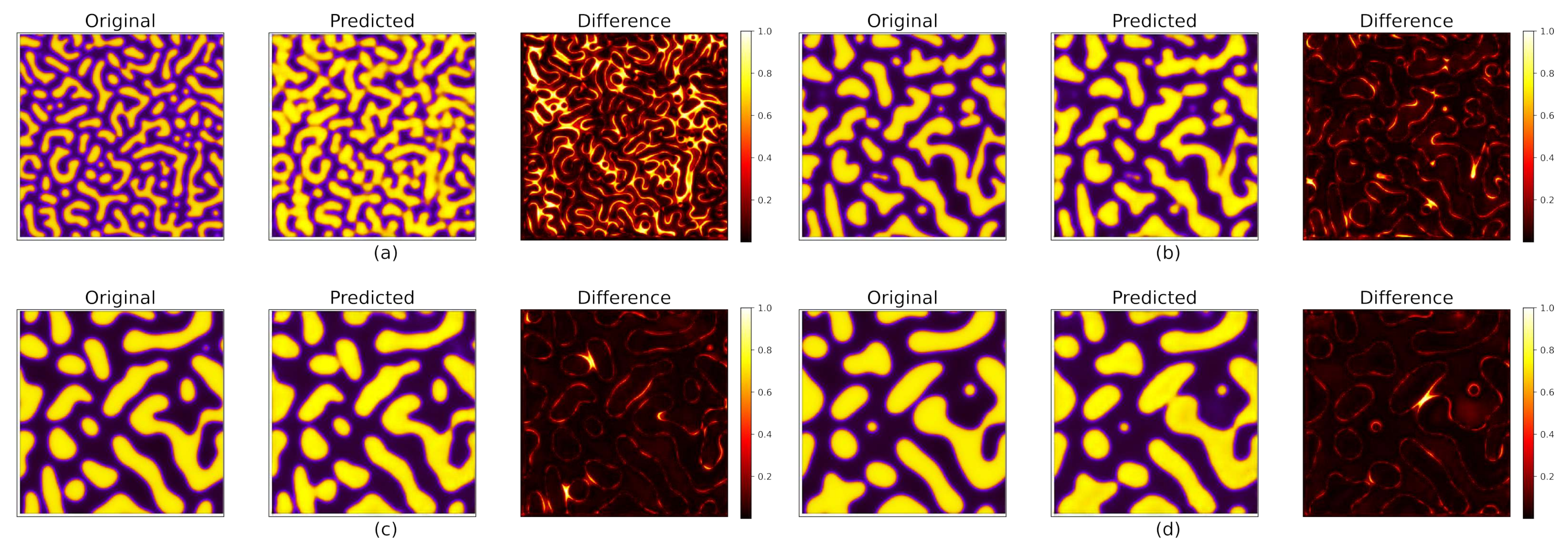}
\caption{A model trained on 40 previous frames to predict the next frame is used to make the final predictions. This figure illustrates the difference between the actual and predicted frame at the 120$^{th}$, 320$^{th}$, 520$^{th}$, and 820$^{th}$ time step as a heatmap.}
\label{fig6}
\end{figure*}

The final optimization step for the ConvLSTM model involves varying the number of previous frames used for predicting the next frame. The model is trained to predict the next frame based on the previous 5, 10, 15, 20, 25, 30, 35, and 40 frames. The predicted frame from ConvLSTM is in the latent space and is projected to its actual dimension with the decoder. As explained previously, the decoder is the second half of the autoencoder, the same model we optimized earlier for dimensionality reduction. After reconstructing the predicted image with the decoder, we compare it with the actual microstructure (directly obtained from phase-field simulations). Panel (a) of Fig.~\ref{fig5} compares the error in the 20$^{th}$ predicted frame based on 5, 10, up to 40 preceding frames prior to the 100$^{th}$ time step; it is evident that as the number of frames grows, the inaccuracy in the predicted frame reduces. Additionally, we find that after 25 frames, the error in the predicted frame does not vary significantly. Similar conclusions can be drawn while predicting the 300$^{th}$, 500$^{th}$, and 800$^{th}$ time steps, as shown in panels (b), (c), and (d) of Fig.~\ref{fig5}.  Thus, one can conclude that the absolute relative error produced in the predicted frame decreases as the number of previous frames used for prediction increases; and more than 25 frames would be a safe choice for the method to predict the $20^{th}$ frame with reasonable accuracy. For the rest of the discussion, we are going to consider microstructures predicted with 40 preceding frames. Fig.~\ref{fig5} also illustrates the error in predicted frames at different stages of the microstructure evolution. Comparing the four panels of Fig.~\ref{fig5}, one can further conclude that error in predicted frames is smaller in the later stages of the microstructure evolution than initial stages. 

\section{Discussion}
\label{discussion}
So far, we have discussed a method that starts with microstructure generation using a Phase-field model. Next, we use the autoencoder, which has two parts. The encoder part carries out the dimensionality reduction of the microstructures to a latent space, and the decoder part reconstructs the original microstructure back from the latent space. Using the data in the latent space (obtained from the encoder), we train a ConvLSTM model, which predicts spatiotemporal evolution in the latent space itself. Finally, using the predicted data in the latent space, the decoder reconstructs the predicted microstructure. 

While the principle is straightforward, one must still optimize the parameters to obtain the highest accuracy at a minimal computation cost. We need to optimize at two levels; first, while training the autoencoder, and next, while training the ConvLSTM.  For example, 2000 images are sufficient to train the autoencoder, and a three-cell encoder offers maximum dimensionality reduction at the expense of minimum data loss (see Fig.~\ref{fig3}). Next, ConvLSTM, trained with four cells and up to 100 epochs, is sufficient for prediction (see Fig.~\ref{fig4}). Since the model is trained to predict the next frame based on the previous frames, optimizing the number of previous frames used for prediction is also essential. We find that the errors between the actual and predicted frame can be minimized by using around 40 previous frames to predict the $20^{th}$ frame, although around 25 previous frames should be sufficient for this purpose (see Fig.~\ref{fig5}).

Interestingly, the prediction quality also depends on the microstructure evolution stage, evident from heatmaps shown in Fig.~\ref{fig6}. These are generated based on predictions using the previous 40 frames. In each case of 100$^{th}$, 300$^{th}$, 500$^{th}$, and 800$^{th}$ time steps during the microstructure evolution, 40 previous frames are taken, and the subsequent 20 frames are predicted, yielding the 120$^{th}$, 320$^{th}$, 520$^{th}$, 820$^{th}$ frame, respectively. Pixel-by-pixel error heatmap shows that, although the model can capture the overall profile at any stage of microstructure evolution, the error is higher during the initial stage (at 120$^{th}$ time step) and decrease significantly at later stages.  Thus, one needs to be cautious, particularly with the predictions during the initial stage of the microstructure evolution.

\begin{figure*}
\includegraphics[width=2\columnwidth]{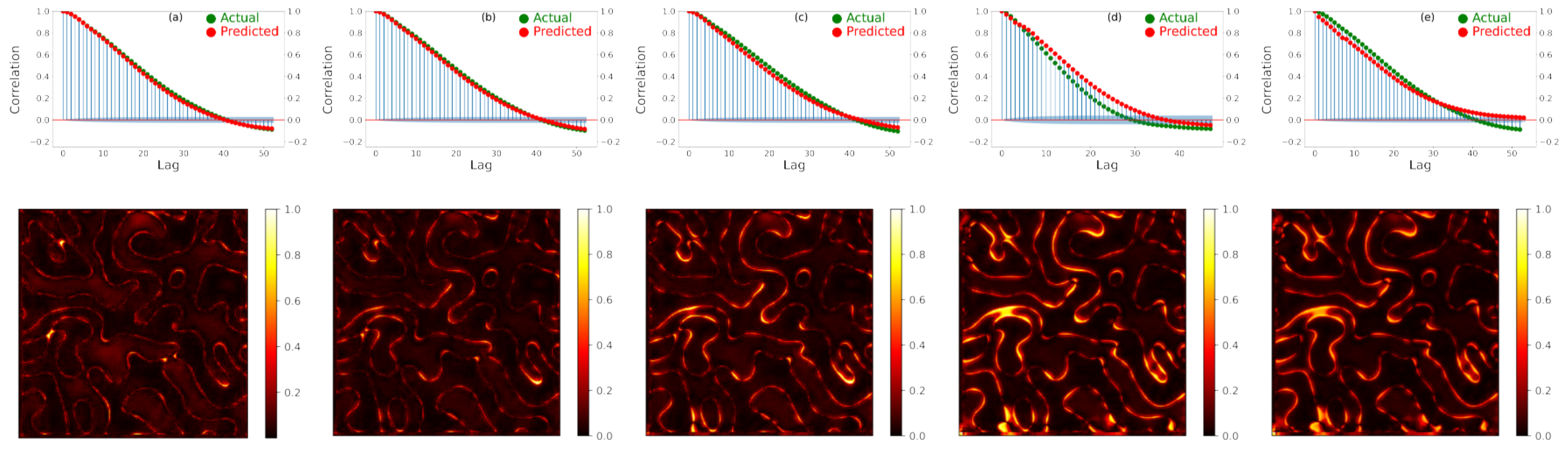}
\caption{Comparative autocorrelation and heatmap of the (a)10$^{th}$, (b)20$^{th}$, (c)30$^{th}$, and (d)40$^{th}$ predicted frame after the model is trained on the preceding 40 frames. The final figure (e) is the 10$^{th}$ frame predicted, using PCA as the dimensionality reduction technique. Comparing (a) and (e) reveals that the autoencoder fares significantly better than PCA.}
\label{fig7}
\end{figure*}

The autoencoder-ConvLSTM model can accelerate in silico study of microstructure evolution, as it can predict $n^{th}$ frame based on previous $m$ frames. Essentially, we are replacing $n$ phase-field steps with machine-learned microstructures. In order to accelerate, one needs to maximize $n$ and minimize $m$, keeping the error within an acceptable limit. For example, if the number of previous frames used is $m=40$, how far can we predict before the errors blow up? A comparison, in terms of the heat map and autocorrelation,~\cite{pfleiderer1993auto} is shown in  Fig.~\ref{fig7} for the $10^{th}$, $20^{th}$, $30^{th}$ and $40^{th}$ frame, predicted from previous 40 frames. Evidently, the predicted spatial domain starts to differ from the actual state as we move further ahead in the time domain. Fig.~\ref{fig7} also presents errors in a predicted microstructure, where PCA is used for the dimensionality reduction instead of the autoencoder. Even the $10^{th}$ predicted frame has unacceptably high errors.

The autoencoder has proven to be a more robust technique than PCA to learn the spatiotemporal variation of microstructure with a minimal dataset for the binary phase. Principal component analysis presumes the linear embedding of microstructure in higher dimensional space. The binary phase microstructure evolution exhibits a non-linear trajectory that PCA failed to capture. Some researchers have already shown the advantage of non-linear dimensionality-reduction techniques such as isometric feature mapping or Isomap, ~\cite{doi:10.1126/science.290.5500.2319} uniform manifold approximation and projection or UMAP.~\cite{mcinnes2020umap}

\section{Conclusions and Future Scope}
\label{conclusion}
In conclusion, the autoencoder-ConvLSTM model provides an accelerated framework for microstructure evolution predictions. The performance depends on two parts of the model. The first part is the autoencoder model, which efficiently reduces dimensions to a compact dataset. The computational cost of the autoencoder is relatively inexpensive. The autoencoder can be trained for as small as $1000$ images for the microstructure evolution during spinodal decomposition in a binary system. It performs very well, both in terms of moderate computational resource requirements and relatively less time taken to reconstruct the image in the original spatial dimension. The second part is the ability of the ConvLSTM neural network to learn spatial information with its temporal dependencies, which can predict the microstructure evolution. The model can be further implemented for more complicated cases, e.g., multi-phase micorstructure evolution, microstructure evolution under the influence of external magnetic field, strain field, etc. 

Recently, several researchers studied distinct methods for utilizing the underlying correlations between datasets derived from diverse data sources with varying accuracy and obtained optimal predictions.~\cite{lu2022multifidelity} One can consider implementing the proposed method's multi-fidelity implementation by including experimental data derived from processes with similar characteristics. In this perspective, data derived from phase-field models using a numerical solver can be regarded as a low-fidelity dataset, while the experimental microstructures and imaging data from analogous processes can be termed high-fidelity datasets.

\section{Acknowledgements}
The authors acknowledge the National Supercomputing Mission (NSM) for providing computing resources of Param Sanganak at IIT Kanpur, which is implemented by C-DAC and supported by the Ministry of Electronics and Information Technology (MeitY) and Department of Science and Technology (DST), Government of India. The authors are also thankful for the HPC facility provided by Computer Center at IIT Kanpur. RM  and SB are thankful for financial support received from Center for Development of Advanced Computing (C-DAC) Project No. Meity/R\&D/HPC/2(1)/2014.

\bibliography{main}
\end{document}